# Semileptonic Decays of $D$ and $B$ Mesons*

S. Güsken[a], K. Schilling[a,b] , G. Siegert[b]

[a] Physics Department, University of Wuppertal, D-42097 Wuppertal, Germany
[b] HRLZ, c/o KFA, D-52425 Jülich, Germany

We report results of our ongoing investigation concerning semileptonic decays of heavy pseudoscalar mesons into pseudoscalar and vector mesons. Particular attention is paid to uncertainties in the $q^2$ and the heavy quark mass dependence of formfactors. Moreover we present a non-perturbative test to the LMK current renormalization scheme for vector current transition matrix elements and find remarkable agreement.

## 1. INTRODUCTION

The accurate determination of Kobayashi Maskawa (KM) matrix elements involving heavy quarks requires control over the low energy QCD parts of the corresponding weak transition amplitudes. Although lattice QCD is the per se method to tackle this regime, it suffers from fact that the lattice resolution is still too poor to represent $b$ quarks directly on the lattice.

With our current high statistics project we push forward to a lattice resolution of $a^{-1} \simeq 3.2$ GeV. This allows for direct calculation of transition amplitudes up to a mass region of $(1-1.6) \times m_D$, keeping discretization errors small. As direct access to the $b$ mass is still excluded, we work with four heavy quark masses, which provide us with a sufficient lever arm for the extrapolation to the $B$ meson.

## 2. LATTICE SETUP

We are working on $24^3 \times 64$ lattice at $\beta = 6.3$, with standard Wilson quarks in the quenched approximation. The heavy quark masses are represented by $\kappa_h = 0.1200, 0.1300, 0.1350, 0.1400$. This covers the physical region $0.8m_c \leq m_h \leq 1.6m_c$. For the light quark masses we have chosen $\kappa_l = 0.1450, 0.1490, 0.1507, 0.1511$, which corresponds to $0.8m_s \leq m_l \leq 3m_s$. The decay amplitudes have been constructed with the initial meson at rest and the final meson carrying spatial momenta $pa = \frac{2\pi}{24} \times \{(0,0,0); (1,0,0); (1,1,0); (1,1,1); (2,0,0)\}$ + permutations. In order to improve on the groundstate projection we have used Wuppertal "gaussian" smeared wavefunctions for the quark fields. Details can be found in ref.[?].

The simulation has been performed on the 32 node CM5 located at the IAI in Wuppertal. We have created a total of 100 independent gauge configurations, which served for calculation of 2- and 3-point correlators.

The analysis has been done so far on a subset of 60 configurations. Using this data set we find $\kappa_{crit} = 0.151818(37)$. The lattice spacing has been determined with three different methods. With R. Sommer's [?] method we find $a_{R_0}^{-1} = 3.314(29)$ GeV, extrapolation of the vector meson mass to $\kappa_{crit}$ yields $a_\rho^{-1} = 3.453(217)$ GeV, and with the stringtension $\sqrt{\sigma} = 440$ MeV we get $a_\sigma^{-1} = 3.210(52)$ GeV. The results agree nicely within errors. In the following we will use $a_{R_0}$ to set the scale.

## 3. $q^2$ DEPENDENCE

The KM matrix elements can be extracted from experimental measurements of the decay widths $\Gamma$ and the branching ratios $R$. In doing so one needs the formfactors[2] $F(q^2)$ , which parametrize the transition matrix elements[?]. As the major contribution to the phase space integrals for $\Gamma$ and $R$ comes from the region of small $q^2$, one con-veniently quotes the values $F(q^2=0)$. Of course knowledge of $F(q^2)$ over the entire $q^2$ region is neccessary in order to calculate $R$ and $\Gamma$ accurately.

*Work supported by DFG grant Schi 257/1-4, Schi 257/3-2, EC contract CHRX-CT92-0051 and Mu 810/3.

[2]For the time being $F(q^2)$ denotes generically the formfactors $f_+, f_0, V, A_1, A_2$.

According to our setup of spatial momenta, the calculation provides us with (up to) 5 entries $F(q^2)$ for each mass combination, which we use to extrapolate to $F(0)$ as well as to learn about the functional dependence of $F(q^2)$.

In our analysis we have tested three different methods to fit our data: Two of them are guided by the pole dominance hypothesis

$$F(q^2) = \frac{F(0)}{1 - \frac{q^2}{m_t^2}} , \qquad (1)$$

where we a) took a fixed pole mass $m_t$, as extracted from the corresponding 2-point correlators, and b) treated $m_t$ as a free parameter. The third Ansatz assumes linear behaviour

$$F(q^2) = a + bq^2 . \qquad (2)$$

Fig. ?? shows a typical example of our formfactor data for weak transitions of a heavy-light pseudoscalar meson into light-light pseudoscalar and vector mesons ($Hl \to ll$), together with the different fits. It turns out that all Ansätze describe the data reasonably well and lead to compatible results at $q^2 = 0$. Therefore the good news is that we are able to determine the value of $F(q^2=0)$ to come out independent of the formfactor model.

On the other hand, however, our data does not clearly distinguish between the different functional forms. This is due to the statistical noise of the data and gives rise to systematic uncertainties.

A similar situation is found for the decays of a heavy-light pseudoscalar meson into heavy-light pseudoscalar and vector mesons ($Hl \to H'l$). Again, the results at $q^2 = 0$ are almost unaltered by the fit method, but sizeable uncertainties concerning the functional form of $F(q^2)$ over the entire $q^2$ range remain.

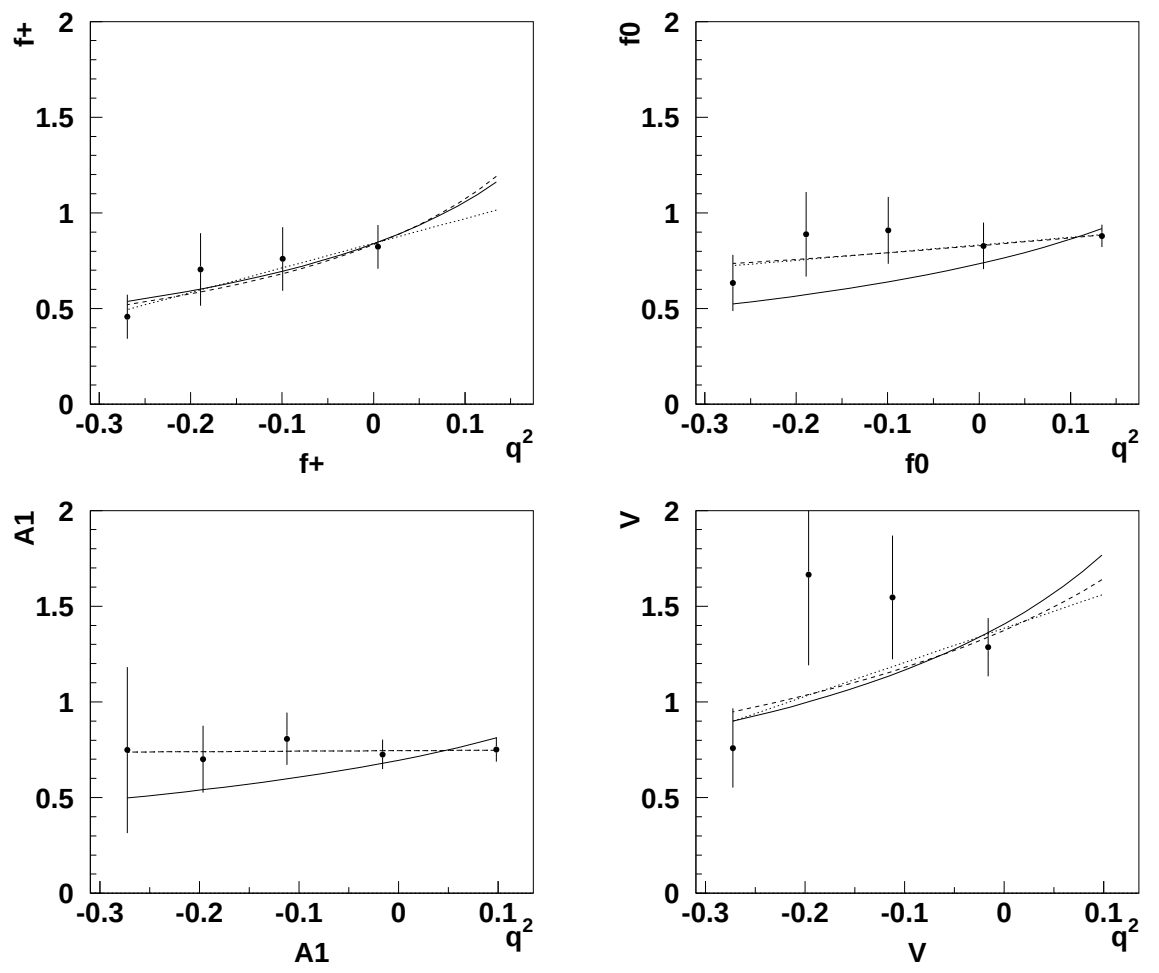

Figure 1. *$q^2$ dependence of formfactors for Hl $\to$ ll transitions. The mass of the heavy initial quark corresponds to $\kappa_i = 0.1350$ and the masses of light final and the light spectator quark is given by $\kappa_l = \kappa_{spec} = 0.1490$. The solid line represents the fit to eq.?? with fixed $m_t$, the dashed line keeps $m_t$ free and the dotted line correponds to a fit to eq.??.*

## 4. RENORMALIZATION

In order to convert the above lattice data into continuum results, we need the renormalization constants $Z_V$ and $Z_A$ of the vector and axialvector currents. It has been pointed out by C.W. Bernard [?] that the renormalization constant of the local vector current $Z_V$ depends strongly on the mass of the quarks involved. Following his arguments this undesired mass dependence can be removed if the (standard) $\sqrt{2\kappa}$ normalization of the quark fields is replaced by the LMK[?] prescription.

As a first step to check this issue[3] with our data we study the ratio

$$R_{H'H}(\vec{p}) = \frac{\langle P(m'_H, m_l)|V_0^{cons}|P(m_H, m_l)\rangle_{\vec{p}}}{\langle P(m'_H, m_l)|V_0^{loc}|P(m_H, m_l)\rangle_{\vec{p}}} . \qquad (3)$$

$V_0^{loc(cons)}$ is the 0'th component of the local(conserved) vector current, and $P$ denotes a pseudoscalar eigenstate.

$R_{HH}(\vec{0})$ yields directly the renormalization factor of $V_0^{loc}$, as $V_0^{cons}$ is unaltered in that instance. For the study of decays we are rather interested

[3] see also [?].

in the ratios $R_{Hh}(\vec{0})$ and $R_{HH}(\vec{p})$ however. The LMK predictions for these ratios read

$$R_{HH}(\vec{0}) = \tilde{Z}_V^{loc} \frac{1 - \frac{3}{4}\frac{\kappa_H}{\kappa_{crit}}}{2\kappa_H} \qquad (4)$$

$$R_{hH}(\vec{0}) = \frac{\tilde{Z}_V^{loc}}{\tilde{Z}_V^{cons}} \frac{1 - \frac{3}{4}\frac{\kappa_H}{\kappa_{crit}}}{2\kappa_H} \qquad (5)$$

$$R_{hH}(\vec{p}) = R_{hH}(\vec{0}) \ . \qquad (6)$$

Taking the standard normalization one gets

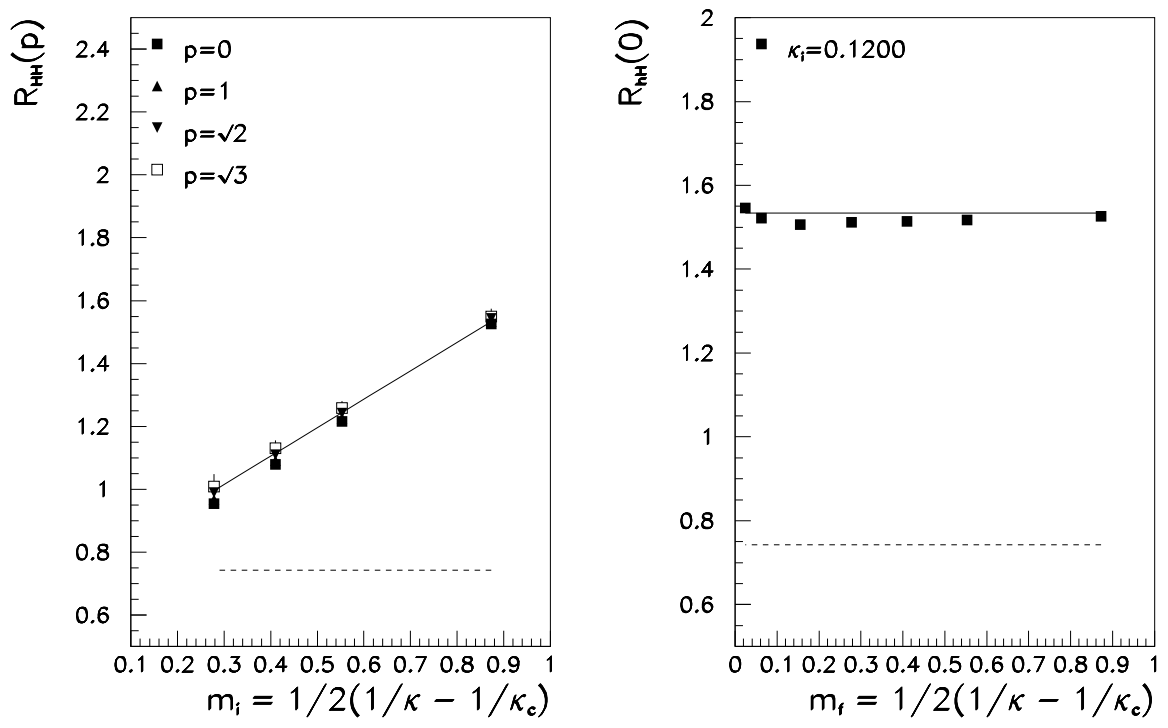

Figure 2. *The ratios $R_{HH}(\vec{p})$ (left) and $R_{hH}(\vec{0})$ as a function of the initial ($\kappa_i$) and final ($\kappa_f$) quark-mass. The solid line represents the LMK prediction, the dashed line corresponds to the standard scheme.*

$R_{hH} = R_{HH} = Z_V^{loc}$. In fig. ?? we display the measured ratios $R_{HH}(\vec{p})$ and $R_{hH}(\vec{0})$ together with the LMK and standard predictions. We have inserted $\tilde{Z}_V^{loc} = 1 - \frac{0.82}{4\pi} g^2$, $Z_V^{loc} = 1 - 0.174081 g^2$, and $\tilde{Z}_V^{cons} = 1$, using the boosted coupling $g^{-2}(\frac{\pi}{a}) = g_0^{-2}\langle 1/3 Tr P_{\mu\nu}\rangle + 0.02461$. The LMK prediction describes the data remarkably well within the entire mass range, whereas the standard prediction fails drastically. In the following we will therefore apply the LMK scheme to our formfactor results, although a similar ratio test for the spacelike components is still missing.

## 5. MASS DEPENDENCE

The determination of the formfactors for $B$ meson decays requires extrapolation in the heavy meson mass. As the exact mass dependence of the formfactors is still unknown we have used several Ansätze for the extrapolation, in order to check on systematic uncertainties:

$$F(m_{PS}, q^2 = 0) = \begin{cases} a) & a + \frac{b}{m_{PS}} \\ b) & \frac{1}{\sqrt{m_{PS}}}(a + \frac{b}{m_{PS}}) \\ c) & \sqrt{m_{PS}}(a + \frac{b}{m_{PS}}) \ . \end{cases} \qquad (7)$$

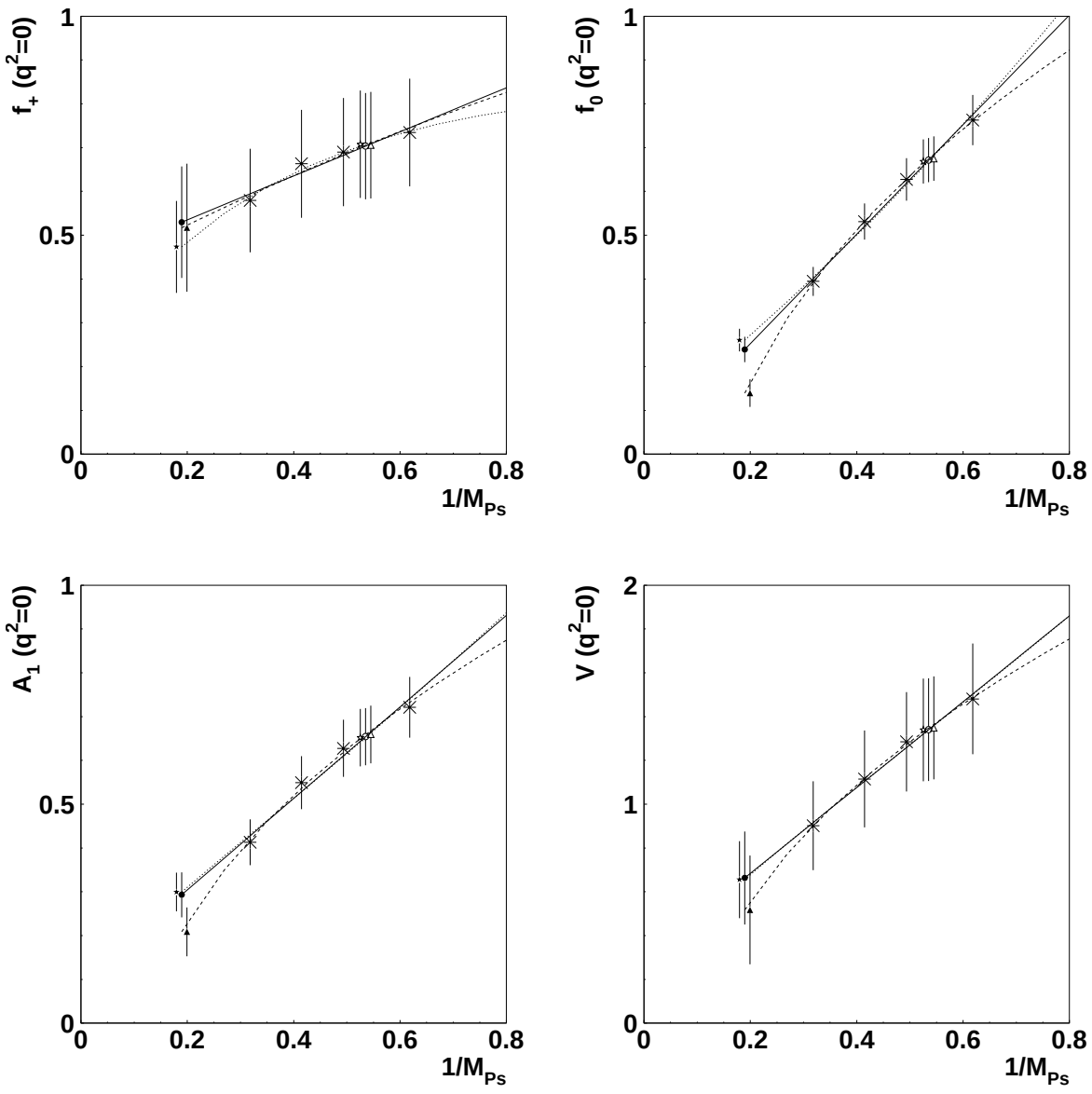

Figure 3. *Mass dependence of formfactors at $q^2 = 0$. Crosses denote the data. The solid line refers to fit method a), the dashed line to b) and the dotted line to c). The extrapolated results are depicted at $1/m_B$ as filled circles (method a), upper triangles (method b), and lower triangles (method c).*

In fig.?? we display the formfactors $F(q^2 = 0)$ for $Hl \to ll$ transitions, as a function of $1/m_{PS}$. Obviously the Ansätze (eq.'s ??) lead to consistent results at $m_{PS} = m_B$.

We mention that the transitions $Hl \to H'l$ exhibit a much weaker mass dependence, and again the results at the $B$ mass are unaltered by the fit method.

Table 1
*Formfactors at $q^2=0$ for weak decays into pseudoscalar mesons.*

| mode | $f_+(0)$ | $f_0(0)$ |
|---|---|---|
| $D \to \pi$ | $0.68(13)^{+10}_{-7}$ | $0.65(5)^{+9}_{-7}$ |
| $D \to K$ | $0.71(12)^{+10}_{-7}$ | $0.67(6)^{+9}_{-7}$ |
| $B \to \pi$ | $0.50(14)^{+7}_{-5}$ | $0.20(3)^{+2}_{-3}$ |
| $B \to D$ | 1.01(60) | 0.88(29) |
| $B_s \to D_s$ | 0.99(42) | 0.85(24) |

Table 2
*Formfactors at $q^2=0$ for weak decays into vector mesons.*

| mode | $V(0)$ | $A_1(0)$ | $A_2(0)$ |
|---|---|---|---|
| $D \to \rho$ | $1.31(25)^{+18}_{-13}$ | $0.59(7)^{+8}_{-6}$ | $0.83(20)^{+12}_{-8}$ |
| $D \to K^\star$ | $1.34(24)^{+19}_{-14}$ | $0.61(6)^{+9}_{-7}$ | $0.83(20)^{+12}_{-8}$ |
| $B \to \rho$ | $0.61(23)^{+9}_{-6}$ | $0.16(4)^{+22}_{-16}$ | $0.72(35)^{+10}_{-7}$ |
| $B \to D^\star$ | 1.84(1.10) | 0.97(32) | 2.76(1.58) |
| $B_s \to D_s^\star$ | 1.84(1.10) | 0.91(24) | 2.25(1.12) |

## 6. RESULTS

In tables ?? and ?? we quote our results for the formfactors at $q^2 = 0$, which have been obtained after extra- and interpolation of the heavy quark mass to the $b$ and $c$ quark and extra- and interpolation of the light quark mass to $m_q = 0$ and $m_q = m_{strange}$ respectively. The errors include both statistical (in brackets) and systematic uncertainties[4]. The results for $D$ decays are in good agreement with experiment and other lattice calculations (see e.g. [?]), giving confidence to our analysis method. For $B \to \pi, \rho$ decays however, the situation is much less settled. The data obey the condition $f_+(0) = f_0(0)$ on a $2\sigma$ level only, indicating systematic uncertainties arising from the extrapolation in the heavy mass. A careful study of this issue will be included in the analysis with full statistics.

[4] This being an intermediate analysis, we have omitted in some cases to quote systematic uncertainties.